# Surface Magnetoelectric Effect in Ferromagnetic Metal Films

Chun-Gang Duan,[1] Julian P. Velev,[2,3] R. F. Sabirianov,[3,4] Ziqiang Zhu,[1] Junhao Chu,[1]

S. S. Jaswal,[2,3] and E. Y. Tsymbal[2,3]

[1]*Key Laboratory of Polarized Materials and Devices, Ministry of Education, East China Normal University, Shanghai 200062, China*

[2]*Department of Physics and Astronomy, University of Nebraska, Lincoln, Nebraska 68588, USA*

[3]*Nebraska Center for Materials and Nanoscience, University of Nebraska, Lincoln, Nebraska 68588, USA*

[4]*Department of Physics, University of Nebraska, Omaha, Nebraska 68182, USA*

A surface magnetoelectric effect is revealed by density-functional calculations that are applied to ferromagnetic Fe(001), Ni(001) and Co(0001) films in the presence of external electric field. The effect originates from spin-dependent screening of the electric field which leads to notable changes in the surface magnetization and the surface magnetocrystalline anisotropy. These results are of considerable interest in the area of electrically-controlled magnetism and magnetoelectric phenomena.

The coupling between ferroelectric and ferromagnetic order parameters in thin-film heterostructures is an exciting new frontier in nanoscale science.[1-5] The underlying physical phenomenon controlling properties of such materials is the *magnetoelectric (ME) effect*[6,7] that determines the induction of magnetization by an electric field or electric polarization by a magnetic field. The recent interest in ME materials is stimulated by their significant technological potential. A prominent example is the control of the ferromagnetic state by an electric field.[8,9] This phenomenon could yield entirely new device concepts, such as electric field-controlled magnetic data storage. In a broader vision, ME effects include not only the coupling between the electric and magnetic polarizations, but also related phenomena such as an electrically-controlled exchange bias,[10,11] and magnetocrystalline anisotropy,[12-14] and the effect of ferroelectricity on spin transport.[15-19]

Two mechanisms for the ME effect have been established: (1) in single-phase ME materials (including multiferroics) an external electric field displaces ions from equilibrium positions which changes the magnetostatic and exchange interactions affecting the magnetization;[20] (2) in composite multiferroic materials, piezoelectric strain in the ferroelectric constituent of the multiferroic heterostructure induces changes in the magnetic properties of the ferromagnetic constituent due to magnetostriction.[9,13,14] Recently, two additional mechanisms for magnetoelectricity have been proposed theoretically: (1) in a heterostructure comprising a ferroelectric insulator and a magnetic material, ferroelectric displacements of atoms at the interface may be reversed by an external electric field resulting in the sizable change of the interface magnetic moment[21] and surface (interface) magnetic anisotropy;[22] (2) in the insulator/ferromagnetic heterostructure, an external electric field polarizes the insulator resulting in the carrier-mediated interface magnetoelectricity.[23]

In this study we explore the ME effect due to the direct influence of an external electric field on magnetic properties of ferromagnetic Fe(001), Ni(001) and Co(0001) films. We show that spin-dependent screening of the electric field leads to spin imbalance of the excess surface charge resulting in notable changes in the surface magnetization and the surface magnetocrystalline anisotropy. We argue that the effect may be used to switch the magnetization between in-plane and out-of-plane orientations, thereby signifying the potential of electrically-controlled magnetism.

When a metal film is exposed to an electric field, the conduction electrons screen the electric field over the screening length of the metal. In ferromagnetic metals, the screening charge is spin-dependent due to exchange interactions.[24] The spin dependence of the screening electrons leads to the induced surface magnetization of the ferromagnet, i.e. the ME effect. Since the electric field does not penetrate into the bulk of metals and the induced electric charge is confined to a depth of the order of atomic dimensions from the surface, this ME effect is limited to the metal surface. Therefore we name it a *surface magnetoelectric effect*.

In order to elucidate the surface magnetoelectric effect quantitatively we carry out density-functional calculations on free-standing ferromagnetic films under the influence of a uniform electric field applied perpendicular to the film surface. The studied systems are bcc Fe(001) ($a$ = 2.87 Å), hcp Co(0001) ($a$ = 2.51 Å, $c/a$ =1.622), and fcc Ni(001) ($a$ = 3.52 Å) films with thicknesses ranging from 1 to 15 monolayers (MLs). The calculations are based on the projector augmented wave (PAW) method implemented in the Vienna *Ab-Initio* Simulation Package (VASP)[25] and include spin-orbit interactions. The exchange-correlation potential is treated in the generalized gradient approximation (GGA). We use the energy cut-off of 500 eV for the plane wave expansion of the PAWs and a 10 x 10 x 1

Monkhorst-Pack grid for *k*-point sampling in the self-consistent calculations. All the structural relaxations are performed until the Hellman-Feynman forces on the relaxed atoms become less than 1 meV/Å. The external electric field is introduced by planar dipole layer method.[26]

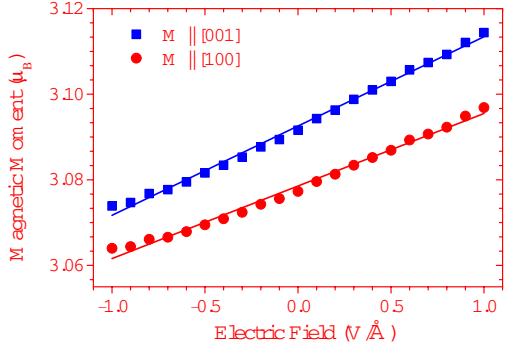

FIG. 1 (color online) Total magnetic moments on the (001) Fe surface as a function of applied electric field for the magnetic moment lying in the plane of the film (along the [100] direction) and perpendicular to the plane (along the [001] direction). The solid lines are a linear fit to the calculated data.

Results of our calculations demonstrate that magnetic moments *do* change under applied electric field. As expected, the magnetoelectric effect is almost solely a surface effect and has little dependence on film thickness. Fig. 1 shows the calculated magnetic moment on the surface Fe atom of a free-standing 15-ML Fe (001) film as a function of an applied electric field for magnetization lying in the plane of the film ([100] direction) and perpendicular to the plane ([001] direction). The spin and orbital contributions to the total moment are of the order of 3.0 and 0.1 $\mu_B$ respectively. It is seen that the magnetic moment changes nearly linearly with the electric field, so that the induced *surface* magnetization $\Delta M$ depends on the applied electric field *E* as follows:

$$\mu_0 \Delta M = \alpha_S E, \qquad (1)$$

where $\alpha_S$ denotes the *surface magnetoelectric coefficient*. Here a positive electric field is defined to be pointed away from the metal film surface. Therefore results for both positive and negative electric fields are obtained at two surfaces within one simulation. From the linear fit to the calculated data shown in Fig. 1 we find that for magnetization in the plane $\alpha_S^{100} \approx 2.4 \times 10^{-14}$ Gcm$^2$/V and for magnetization perpendicular to the plane $\alpha_S^{001} \approx 2.9 \times 10^{-14}$ Gcm$^2$/V. The vertical separation between the two lines in Fig. 1 for a given applied electric field is the difference in the orbital moments in the two magnetization directions (as shown later in Fig. 4) because the corresponding spin moments are essentially the same.

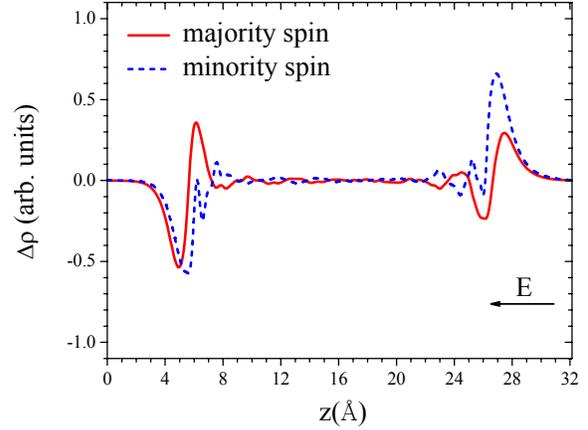

FIG. 2 (color online) Induced *xy*-averaged electron charge densities, $\Delta \rho = \rho(E) - \rho(0)$, along the *z* direction normal to the *xy* film plane for a 21 Å-thick Fe film (located between about 5.5 and 26.5 Å) for majority- (solid line) and minority- (dashed line) spin electrons. The applied external electric field is *E* = 1.0 V/Å, pointing from right to left.

The origin of this ME effect stems from induced spin-dependent charge densities on the surfaces of the film. Fig. 2 displays the differences in the charge densities between the disturbed (*E* = 1.0 V/Å) and undisturbed (*E* = 0) Fe film. It is seen that the surface dipoles are formed to screen the electric field inside the film and the induced charge density is strongly spin-polarized. We note the presence of the Friedel-like oscillation of the change density in Fig. 2, which is typical for the electron screening effect.[27] However, unlike the case of normal metals, the charge oscillations have multiple periods, and the oscillations of the majority- and minority-spin electrons are different. This is because the majority- and minority-spin electrons have different Fermi wave vectors, and they are coupled in the dielectric response.[24]

Due to the spin imbalance of the screening charge, there is an induced surface magnetization reflecting the presence of the surface ME effect. Fig. 3 visualizes the induced spin density across the Fe film. It is clearly seen that the ME effect is confined to the surfaces of the film. The net induced spin densities at the two opposite surfaces have different signs, as the applied electric fields are oppositely oriented with respect to the two surfaces of the film.

It is interesting to compare the calculated magnitude of the surface ME coefficient due to the direct influence of an external electric field with those predicted earlier for BaTiO$_3$/Fe and SrTiO$_3$/SrRuO$_3$ structures. In the later case, the ME effect originates from the capacitive accumulation of spin-polarized carriers at the SrTiO$_3$/SrRuO$_3$ interface under an external electric field. The calculations of ref. 23 suggest that for a 7 unit cell thick SrTiO$_3$ (*a* = 3.904 Å) layer the applied voltage of 27.8 mV induces a net magnetic moment of 2.5×10$^{-3}$ $\mu_B$ per surface unit cell. The corresponding surface ME coefficient is $\alpha_S \approx 2\times 10^{-12}$ Gcm$^2$/V. This value is higher by two orders of magnitude than what we obtained for Fe film. This difference is related to the dielectric constant of the dielectric at the dielectric/metal interface. For a given applied electric field, the screening charge in the metal is proportional to the dielectric constant. According to ref. 17, SrTiO$_3$ has a very large dielectric constant, $\varepsilon \approx 490$, compared to $\varepsilon = 1$ in the present calculations. Therefore, the presence of a dielectric with large dielectric constant can significantly amplify the surface ME effect.

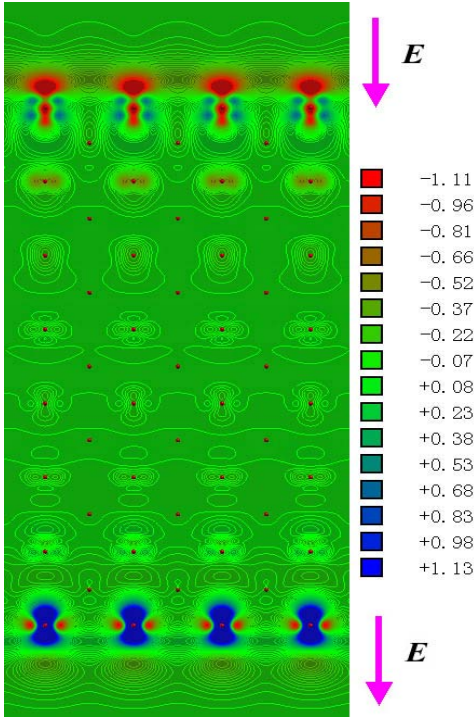

FIG. 3 (color online) Induced spin densities, $\Delta\sigma = \sigma(E) - \sigma(0)$, in arbitrary units for a 15-ML-thick Fe film (about 21 Å) under the influence of an electric field of 1 V/Å.

In the case of BaTiO$_3$/Fe structure, the predicted ME effect occurs as a result of the reversal of the polarization orientation due to an applied electric field and originates from the change in the interface bonding strength which dominates over the screening charge contribution. This ME effect is strongly non-linear and does not follow a simple relation given by Eq. (1). Nevertheless, to have a crude estimate of the surface ME coefficient we assume that the polarization of BaTiO$_3$ can be switched at the coercive field of $E_c = 100$ kV/cm resulting in the change of the interface magnetic moment of more than 0.3 $\mu_B$ per unit surface cell. We find that the surface ME coefficient is then $\alpha_s \approx 2 \times 10^{-10}$ Gcm$^2$/V. This value is much larger than the electric-field-induced effects discussed above.

Calculations performed for magnetic Co (0001) and Ni (001) films confirm qualitatively predictions obtained for Fe. We find that the surface ME coefficients (in units of $10^{-14}$ Gcm$^2$/V) are $\alpha_S^{100} \approx 1.6$, $\alpha_S^{001} \approx 1.7$ for 9-ML Co (0001) film and $\alpha_S^{100} \approx 3.0$, $\alpha_S^{001} \approx 2.4$ for 9-ML Ni (001) film. Note that the predicted surface ME coefficients have the same sign for the three kinds of metal films. This is in disagreement with the prediction of ref. 24 suggesting that Fe film has different sign of the induced magnetic moment from that in Co and Ni films. Obviously a simple free-electron-like model adopted in ref. 24 is unable to take into account all the electronic effects which occur in the 3$d$ metals due to the presence of the exchange-splitting $d$ bands, their hybridizations with $s$ and $p$ bands, and the difference between bulk and surface electronic structures.

The origin of the predicted *positive* ME coefficient for all the ferromagnetic films can be understood within a simple model. The model assumes a localization of the screening charge within the first atomic ML on the metal surface and a rigid shift of the chemical potential on the surface in response to the applied electric field $E$. According to this model the surface charge density $\sigma = \varepsilon E$ is unequally distributed between the surface majority- and minority-spin states resulting in the surface ME coefficient

$$\alpha_s = -\frac{\varepsilon \mu_B}{ec^2} \frac{n^\uparrow - n^\downarrow}{n^\uparrow + n^\downarrow}. \quad (2)$$

Here $n^\uparrow$ and $n^\downarrow$ are majority- and minority-spin surface DOS at the Fermi energy and $\varepsilon$ is the dielectric constant of the dielectric adjacent to the ferromagnetic surface (in our case of vacuum $\varepsilon = 1$). As is seen from Eq. (2), $\alpha_s$ is proportional to the spin polarization of the surface DOS. It is known that at the Fe (001), Co(0001) and Ni(001) surfaces, the minority-spin states are dominant near the Fermi level.[28] Therefore the accumulation of electrons in response to the application of an inward (negative) electric field results in a decrease of the surface spin moment, due to the dominating minority-spin character at the Fermi energy. Using Eq.(2) and the calculated surface DOS, we find $\alpha_S \approx$ 5.2, 4.9 and 5.5 $\times$ 10$^{-14}$ Gcm$^2$/V for Fe (001), Co (0001), and Ni (001) respectively. These values are in a qualitative agreement with the results of our density-functional calculations.

Another important consequence of applying electric field to the ferromagnetic metal films is the change of their magnetocrystalline anisotropy energy (MAE), which may be considered as another manifestation of the ME effect. The MAE, whose physical origin is the spin-orbit coupling,[29] plays an important role in high anisotropy materials used in modern magnetic storage technologies.[30] Recently we have demonstrated that the magnetic anisotropy of the ferromagnetic film can be altered by switching the polarization of the adjacent ferroelectric through applied bias voltage. The effect occurs due to the change of the electronic structure at the interface region, which is produced by ferroelectric displacements and mediated by interface bonding. Here we demonstrate the direct impact of an external electric field on the MAE.

Following our previous study, we decompose the MAE for the whole film into individual contributions from each constituent atom. By doing this, we find that the MAE changes for the ferromagnetic film under various electric fields, again, mainly occur at the surface. To be specific, an outward (positive) electric field enhances and an inward (negative) electric field reduces the individual MAE contribution from surface atoms. As is seen from Fig. 4, the orbital moment anisotropy ($M_L$ [001] - $M_L$[100]) of the surface Fe atoms increases monotonically with the increase of applied electric field. This leads to a significant increase of the surface MAE when the electric field changes from inward (negative) to outward (positive) direction, evident from Fig. 4. In particular, when the electric field of 0.5 V/Å switches from negative to positive, there is a change of about 30% in the surface MAE.

The predicted phenomenon can be used for switching the magnetization by an applied electric field. The total magnetic anisotropy energy per unit area of the film involves the magnetostatic shape anisotropy energy $K_m = -2\pi M^2 t$, where $M$ is the magnetization and $t$ is film thickness. The shape anisotropy favors in-plane alignment of magnetization, whereas positive *MAE* favors out-of-plane alignment. Thus, with the ME control of the surface magnetocrystalline anisotropy and thickness dependent shape anisotropy, it is possible to design ferromagnetic films with the anisotropy switchable between in-plane and out-of-plane orientations.

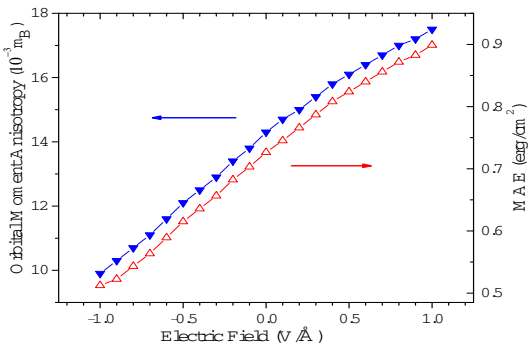

FIG. 4 (color online) Electric field induced changes in calculated orbital moment anisotropy ($\Delta M_L = M_L$ [001] - $M_L$ [100], in units of $10^{-3}\mu_B$) of the surface Fe atom and surface magnetocrystalline anisotropy energy (MAE) for 15 ML thick Fe (001) slab.

The predicted ME effects in ferromagnetic metal films are significant only when the applied field is very large, e.g., of a few hundred mV/Å. However, such strong electric fields could be achieved experimentally using a scanning tunneling microscope tip over the magnetic metal film surface. Since the screening effect can be dramatically enhanced by high-κ dielectrics, the applied electric field could be much smaller to have observable effects on the related devices.

To summarize, our first-principle calculations show that the surface magnetoelectric effect exists in ferromagnetic metal films due to spin-dependent screening of electric field at the metal surfaces. The external electric field induces notable changes in the surface magnetization and the surface magnetocrystalline anisotropy of a ferromagnetic metal. The magnitude and sign of the surface magnetoelectric coefficient depends on the density and spin polarization of the charge carriers near the Fermi level of the ferromagnetic metal film. This purely electric field driven magnetoelectric effect may be interesting for application in advanced magnetic and electronic devices.

The research at East China Normal University was supported by the NSFC (50771072), 973 Program No. 2007CB924900, and Shanghai Basic Research Program No. 07JC14018. The research at University of Nebraska was supported by NSF-MRSEC, the Nanoelectronics Research Initiative, the Office of Naval Research, and the Nebraska Research Initiative. C.-G. D. thanks Nicola Spaldin for stimulating discussions.

1. T. Kimura *et al.*, Nature **426**, 55 (2003).
2. T. Lottermoser *et al.*, Nature **430**, 541 (2004).
3. W. Eerenstein, N. D. Mathur and J. F. Scott, Nature **442**, 759 (2006).
4. R. Ramesh and N. A. Spaldin, Nature Mater. **6**, 21 (2007).
5. C.-W. Nan *et al.*, J. Appl. Phys. **103**, 031101 (2008).
6. M. Fiebig, J. Phys. D **38**, R123 (2005).
7. N. A. Spaldin and M. Fiebig, Science **309**, 391 (2005).
8. T. Zhao *et al.*, Nature Mater. **5**, 823 (2006)
9. F. Zavaliche *et al.*, Nano Lett. **7**, 1586 (2007).
10. P. Borisov *et al.*, Phys. Rev. Lett. **94**, 117203 (2005).
11. V. Laukhin *et al.*, Phys. Rev. Lett. **97**, 227201 (2006).
12. M. Weisheit *et al.*, Science **315**, 349 (2007).
13. W. Eerenstein *et al.*, Nature Mater. **6**, 348 (2007).
14. S. Sahoo *et al.*, Phys. Rev. B **76**, 092108 (2007).
15. E. Y. Tsymbal and H. Kohlstedt, Science **313**, 181 (2006).
16. M. Y. Zhuravlev *et al.*, Phys. Rev. Lett. **94**, 246802 (2005).
17. Ch. Binek and B. Doudin, J. Phys.: Cond. Matt. **17**, L39 (2005).
18. M. Gajek *et al.*, Nature Mater. **6**, 296 (2007).
19. J. P. Velev *et al.*, Phys. Rev. Lett. **98**, 137201 (2007).
20. I. Dzyaloshinskii, Sov. Phys. J. Expt. Theor. Physics. **10**, 628629 (1960).
21. C.-G. Duan, S. S. Jaswal, and E. Y. Tsymbal, Phys. Rev. Lett. **97**, 047201 (2006).
22. C.-G. Duan *et al.*, Appl. Phys. Lett. **92**, 122905 (2008).
23. J. M. Rondinelli, M. Stengel, and N. A. Spaldin, Nature Nanotech. **3**, 46 (2008).
24. S. Zhang, Phys. Rev. Lett. **83**, 640 (1999).
25. G. Kresse and D. Joubert, Phys. Rev. B **59**, 1758 (1999).
26. J. Neugebauer and M. Scheffler, Phys. Rev. B **46**, 16067 (1992).
27. K. M. Indlekofer and H. Kohlstedt, Europhys. Lett. **72**, 282 (2005).
28. O. Hjortstam *et al.*, Phys. Rev. B **53**, 9204 (1996).
29. J. H. Van Vleck, Phys. Rev. **52**, 1178 (1937).
30. R. C. O'Handley, *Modern Magnetic Materials: Principles and Applications* (Wiley-VCH, 1999).